\documentclass{aa}  

\usepackage{graphicx}
\usepackage{natbib}
\bibpunct{(}{)}{;}{a}{}{,} 

\usepackage[varg]{txfonts}
\usepackage{esvect}
\begin{document}
\title{The origin of organic emission in NGC 2071\thanks{This paper uses \textit{Herschel} observations. {\it Herschel} is an ESA space observatory with science instruments provided by European-led Principal Investigator consortia and with important participation from NASA. This paper also makes use of SMA observations. The Sub-millimeter Array is a joint project between the Smithsonian Astrophysical Observatory and the Academia Sinica Institute of Astronomy and Astrophysics and is funded by the Smithsonian Institution and the Academia Sinica.
}}
\author{ 
T.A. van Kempen \inst{1} \and
C. M$^{\rm{c}}$Coey \inst{2} \and
S. Tisi \inst{2} \and
D. Johnstone \inst{3,4,5} \and
M. Fich \inst{2} 
}

\institute{
$^1$Leiden Observatory, Leiden University, Niels Bohrweg 2, 2333 CA, Leiden, the Netherlands \\
\email{kempen@strw.leidenuniv.nl}\\
$^2$Department of Physics and Astronomy, University of Waterloo, Waterloo, Ontario, N2L 3G1, Canada \\ 
$^3$Joint Astronomy Centre, 660 North A'ohoku Place, University Park, Hilo, HI 96720, USA \\
$^4$National Research Council Canada, Herzberg
Institute of Astrophysics, 5071 West Saanich Rd, Victoria, BC, V9E
2E7, Canada\\
$^5$Department of Physics \& Astronomy, University of Victoria,
Victoria, BC, V8P 1A1, Canada\\
}

\date{Received Jan 1, 2014}

\abstract
   {The physical origin behind organic emission lines in embedded low-mass star formation has been 
   fiercely debated in the last two decades. A multitude of scenarios have been proposed, from a hot corino to PDRs on cavity walls to shock excitation. }
   {The aim of this paper is to determine the location and the corresponding physical conditions of the gas responsible for organics emission lines. The outflows around the small protocluster NGC 2071 are an ideal testbed to differentiate between various scenarios.}
   {Using Herschel-HIFI and the SubMillimeter Array, observations of CH$_3$OH, H$_2$CO and CH$_3$CN emission lines over a wide range of excitation energies were obtained. 
   Comparisons to a grid of radiative transfer models provide constraints on the physical conditions. Comparison to H$_2$O line shape is able to trace gas-phase synthesis versus a sputtered origin. 
   }
   {Emission of organics originates in three separate spots: the continuum sources IRS 1 ('B') and IRS 3 ('A') as well as a new outflow position ('F'). Densities are above 10$^7$ cm$^{-3}$ and temperatures between 100 to 200 K.
   CH$_3$OH emission observed with HIFI originates in all three regions and cannot be associated with a single region. Very little organic emission originates outside of these regions.}
   {Although the three regions are small ($<$1,500 AU), gas-phase organics likely originate from sputtering of ices due to outflow activity. The derived high densities ($>$10$^7$ cm$^{-3}$) are likely a requirement for organic molecules to survive from being immediately destroyed by shock products after evaporation. The lack of spatially extended emission confirms that organic molecules cannot (re)form through gas-phase synthesis, as opposed to H$_2$O, which shows strong line wing emission. The lack of CH$_3$CN emission at 'F' is evidence for a different history of ice processing due to the absence of  a protostar at that location and recent ice mantle evaporation.}


\maketitle

%
\def\Msol{M$_{\odot}$}

\def\PlaceFigureCH3CN{
\begin{figure}
\centering
\includegraphics[angle=270,width=8cm]{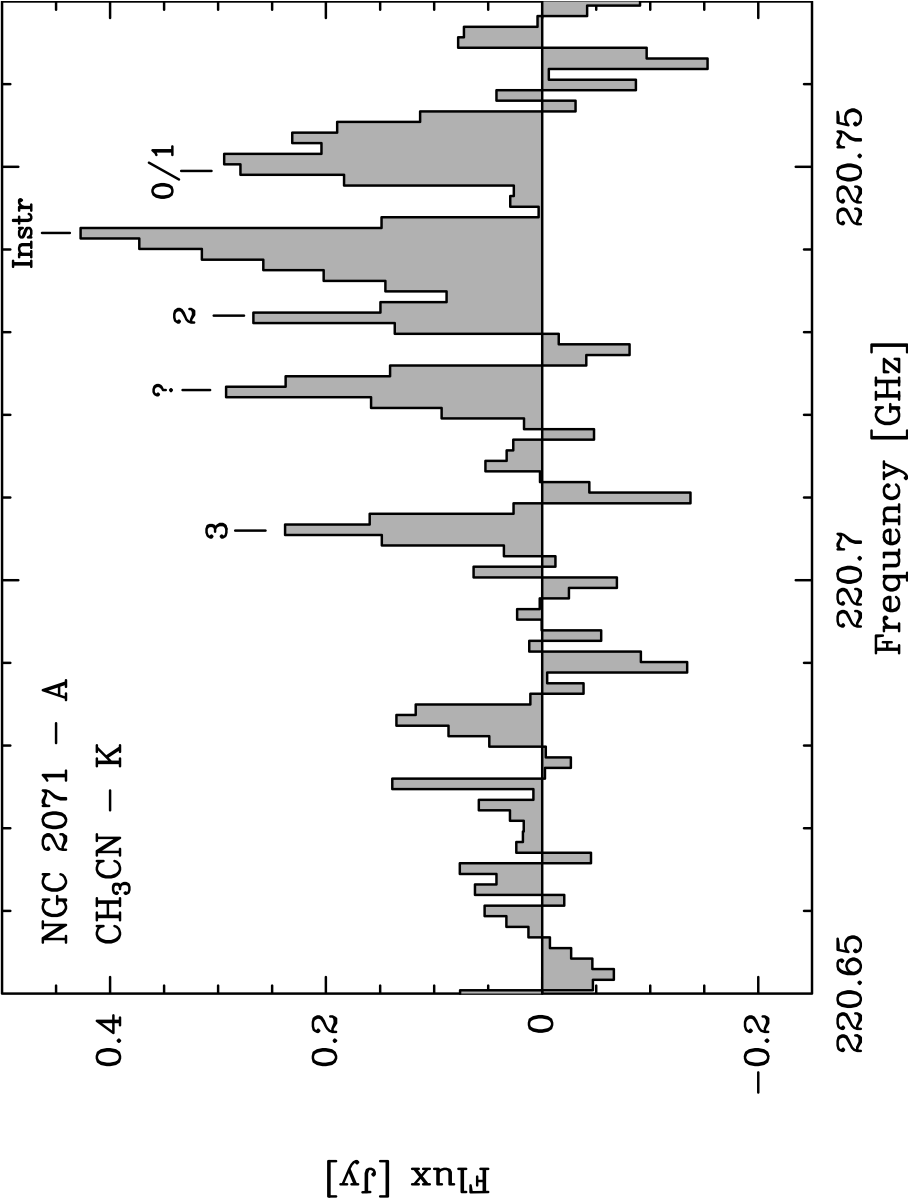}
     \caption{CH$_3$CN $K$ = 12 -- 11 ladder between 220.65 and 220.8 GHz observed by the SMA at the position of region 'A'. The instrumental feature between the 2 and 0/1 blended transitions is labelled with $"$Instr$"$. }
     \label{fig:ch3cn}
\end{figure}
}

\def\PlaceFigurehifi{
\begin{figure}
\centering
\includegraphics[width=8cm]{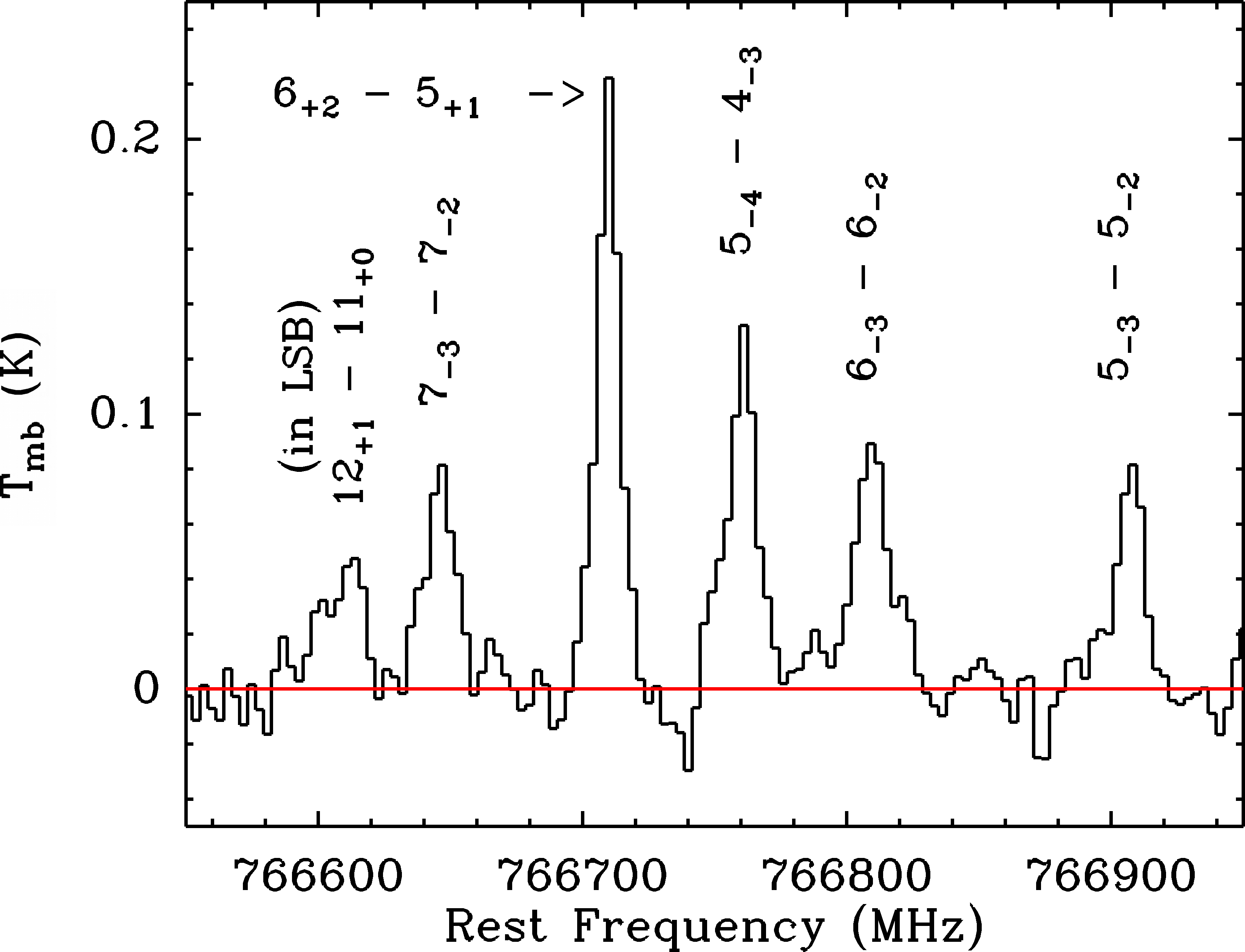}
     \caption{Detected CH$_3$OH lines observed with HIFI in 400 MHz centered at 766.75 GHz. Note that the 12$_{+1}$-11$_{+0}$ line is from the lower sideband, with a rest frequency of 751.55 GHz. Channels are resampled to a width of 0.75 km s$^{-1}$. The baseline is shown as a horizontal red line. At this frequency, the beam of HIFI is 26$''$.}
     \label{fig:ch3ohmethanol}
\end{figure}
}

\def\PlaceFiguremetha{
\begin{figure*}
\centering
\includegraphics[width=5.8cm]{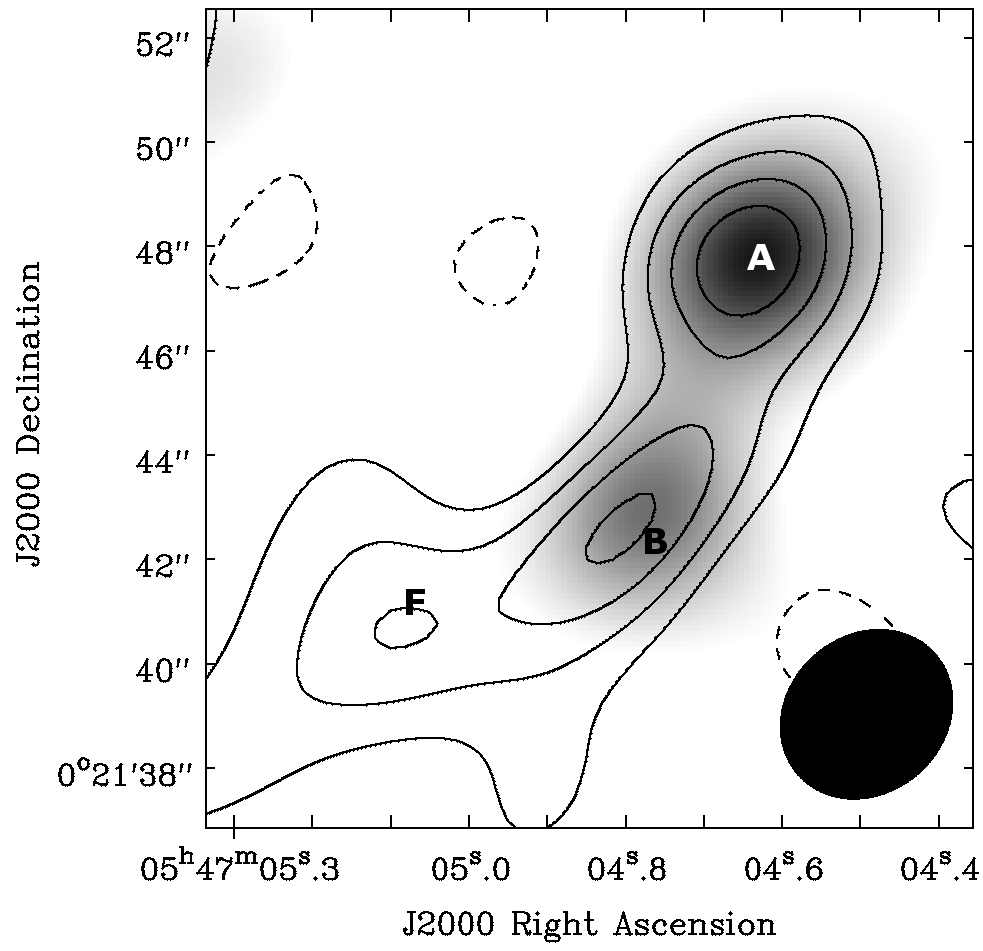}
\includegraphics[width=6cm]{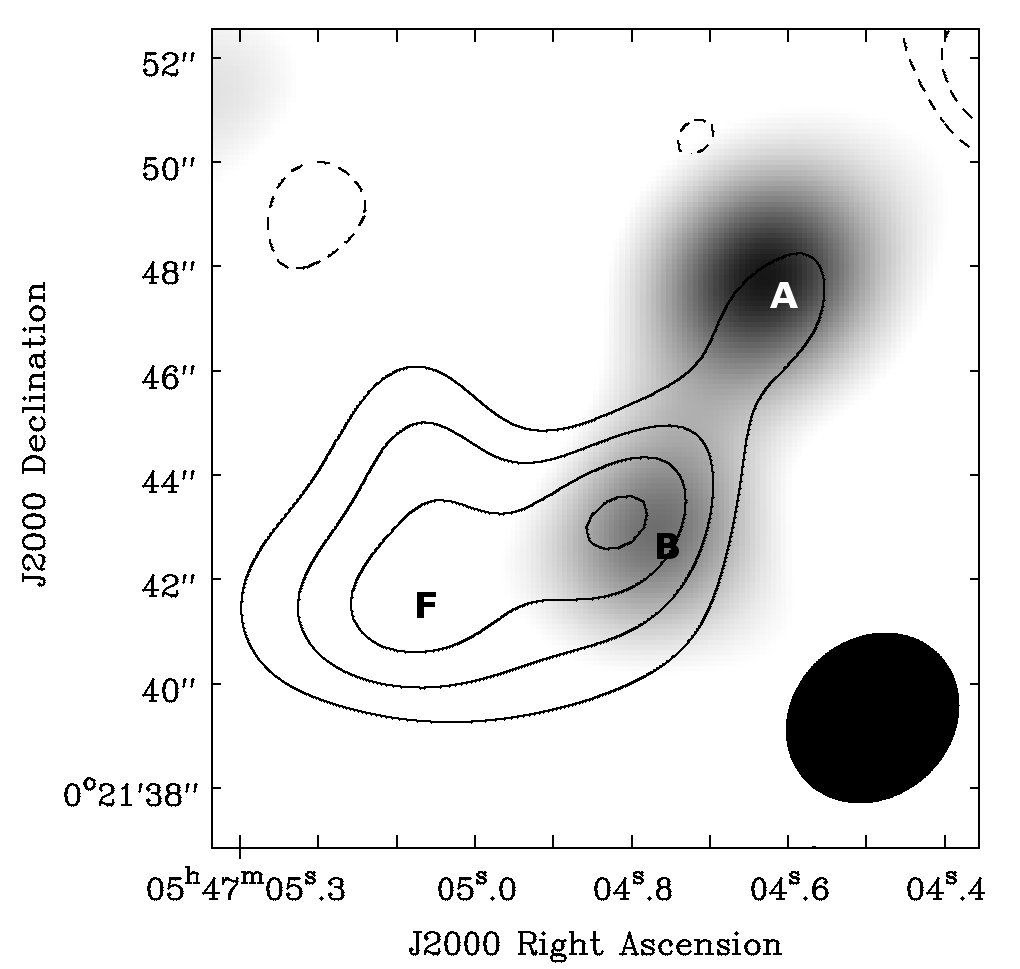}
\includegraphics[width=5.77cm]{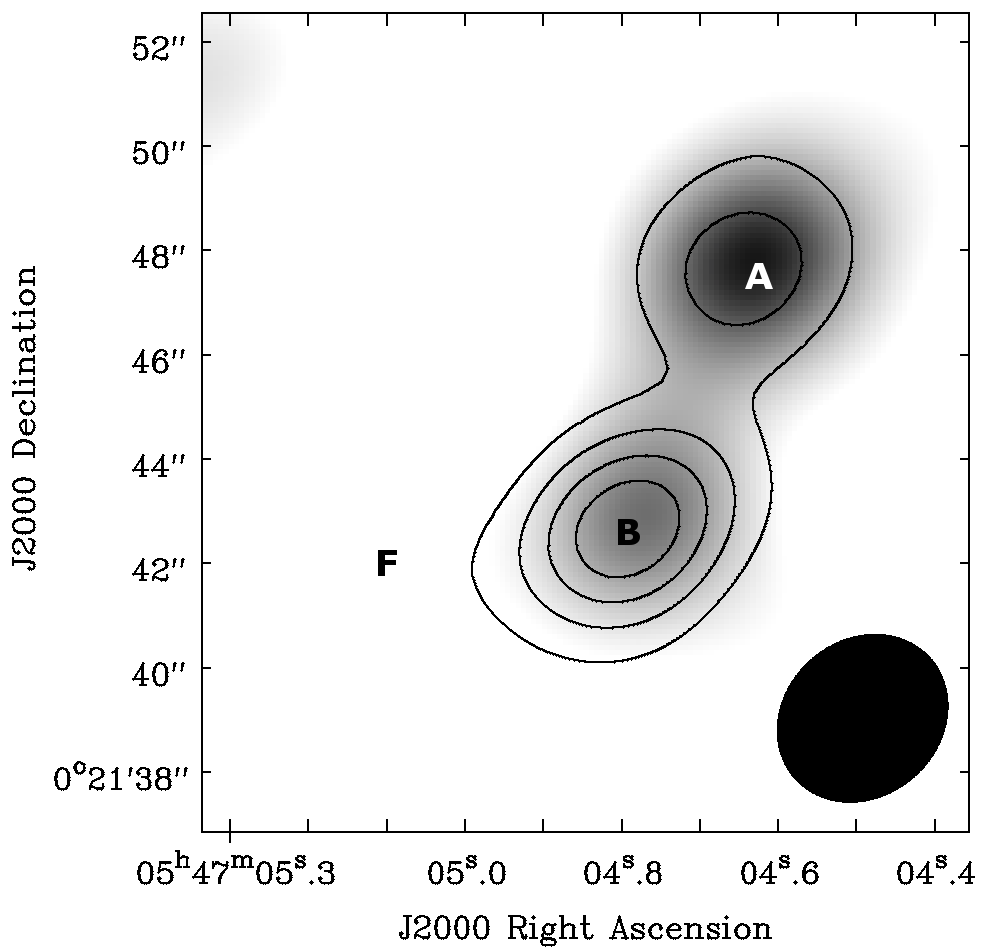}
     \caption{H$_2$CO 218.48 GHz (\textit{left}), CH$_3$OH 218.44 GHz (\textit{middle}) and CH$_3$CN 220.71 GHz (\textit{right}) maps, revealing that the emission originates from the 'A' and 'B' cores and the blue flow (marked A, B and F). Contours are overplotted on the continuum emission from \cite{2012ApJ...751..137V} in 20, 40, 60 and 80$\%$ of the peak line integrated intensity, which is always located at core B.}
     \label{fig:maps}
\end{figure*}
}

\def\PlaceFigureco{
\begin{figure}
\centering
\includegraphics[width=8cm]{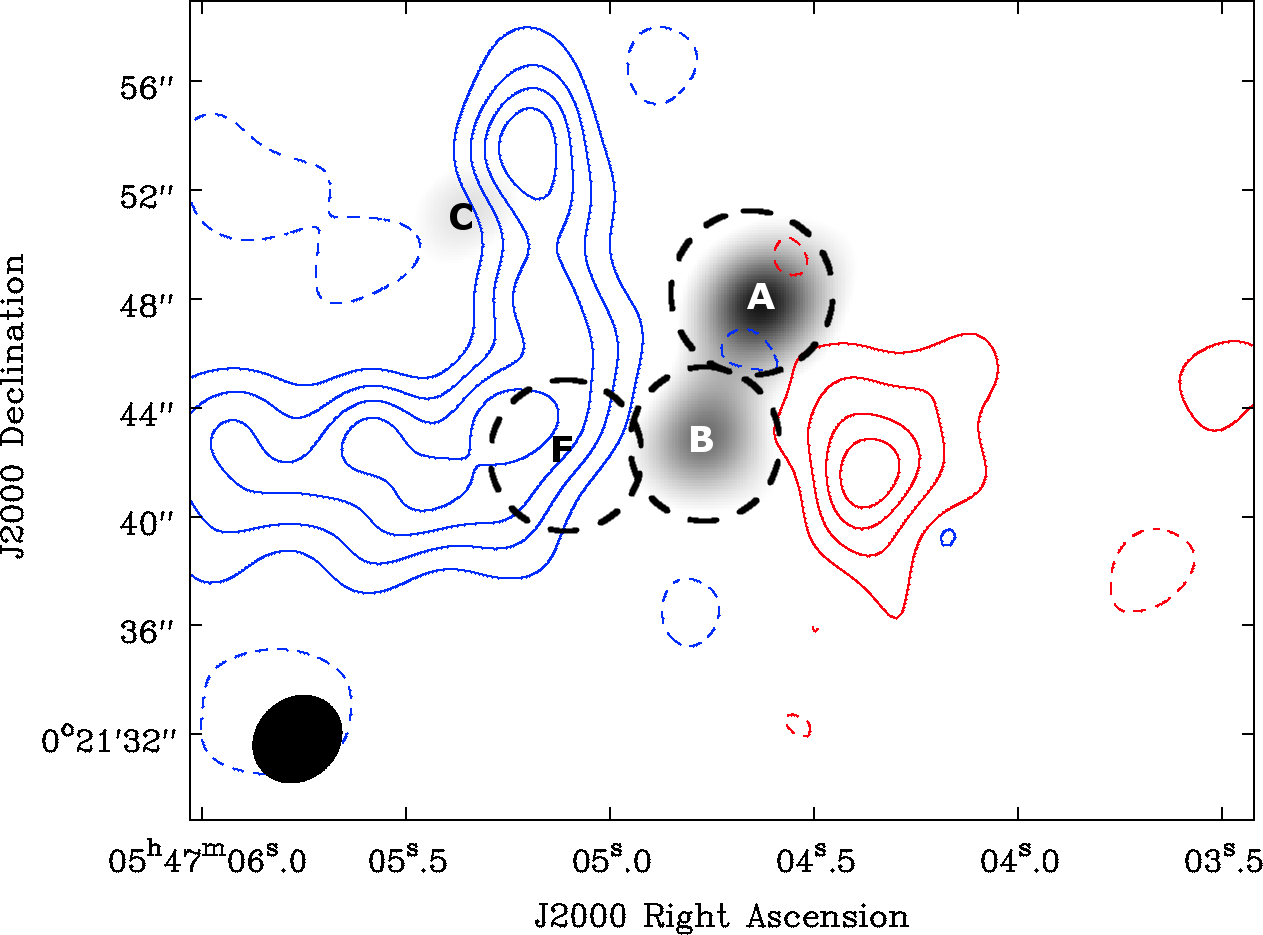}
     \caption{Integrated line wings of $^{13}$CO emission (blue and red) around NGC 2071 obtained with the SMA. It is overplotted on the 1.3 mm continuum emission (greyscale) in steps of 3,6,9,12 $\sigma$ (see van Kempen et al., 2012). Locations of continuum sources (A, B, and C) are labelled. The beam size of 3.4$''\times$2.9$''$ is shown in a black ellipse in the lower left. Dashed circles indicate the regions integrated for organic emission, including region F (see text).}
     \label{fig:13co}
\end{figure}
}

\def\PlaceFiguremetharot{
\begin{figure}
\centering
\includegraphics[width=8cm]{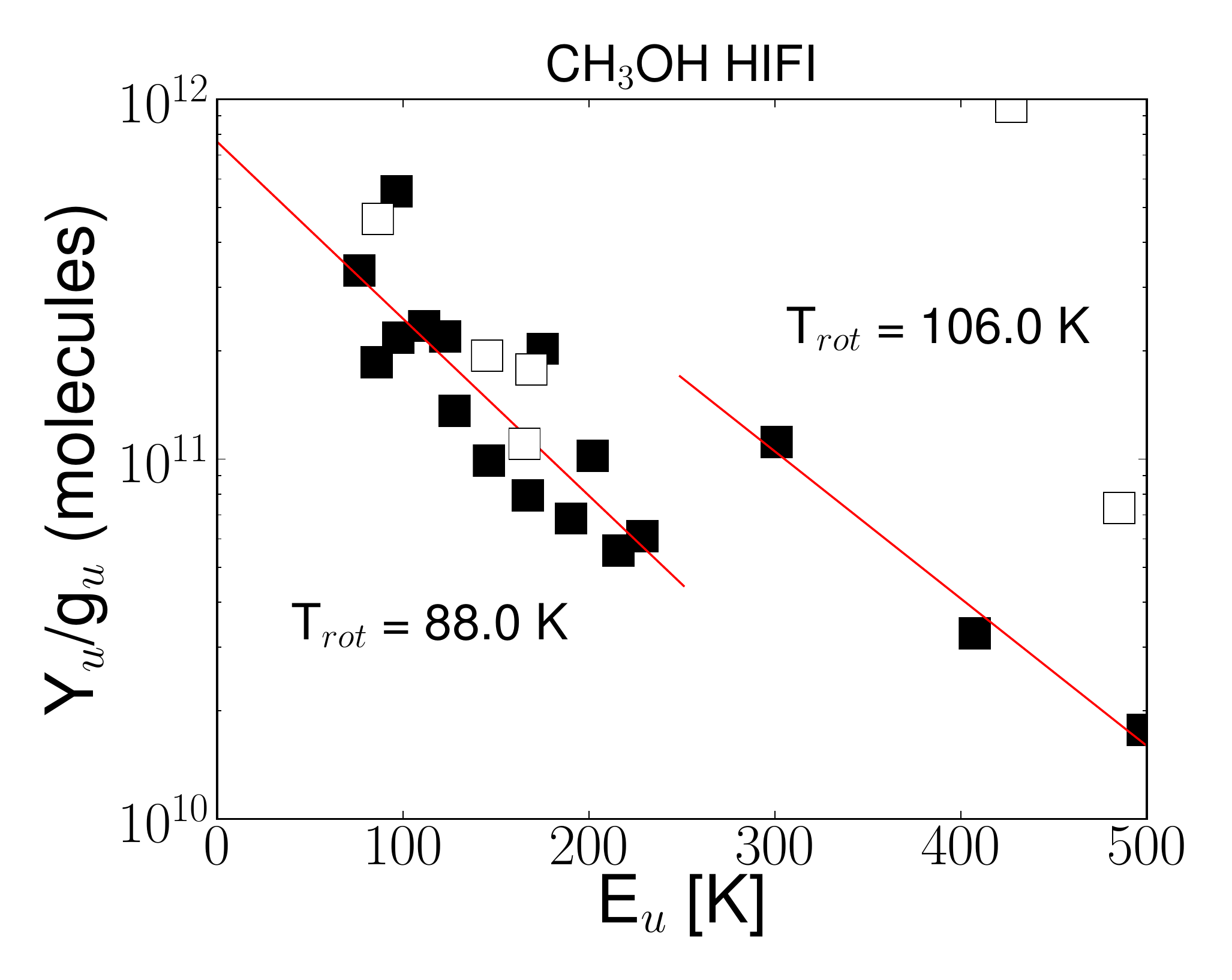}
     \caption{Rotational diagram of the CH$_3$OH in HIFI with the fitted rotational temperatures of both components. Open squares are A-type CH$_3$OH identifications.}
     \label{fig:meth_rot}
\end{figure}
}

\def\PlaceFiguresmaradex{
\begin{figure*}
\centering
\includegraphics[width=6cm]{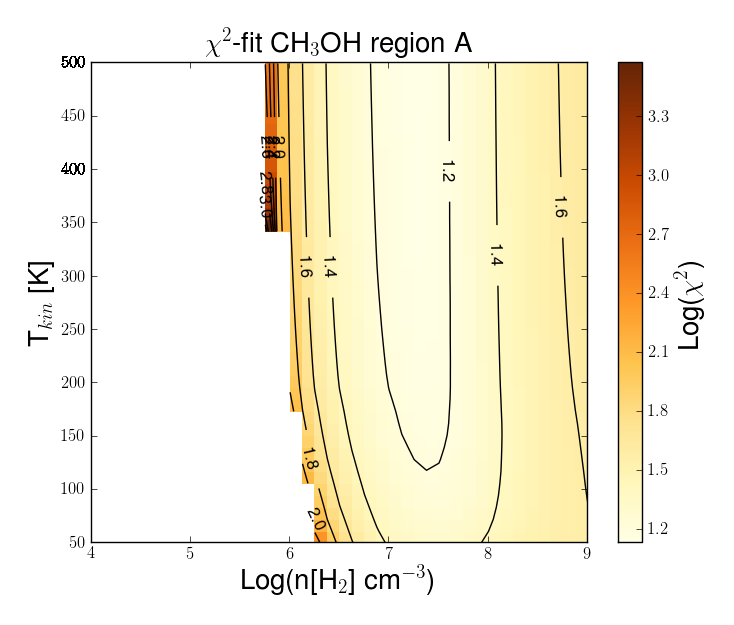}
\includegraphics[width=6cm]{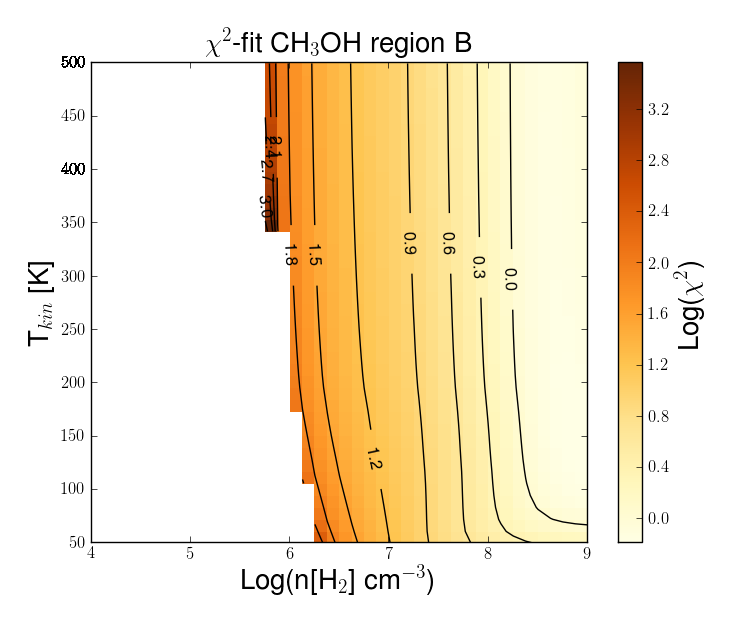}
\includegraphics[width=6cm]{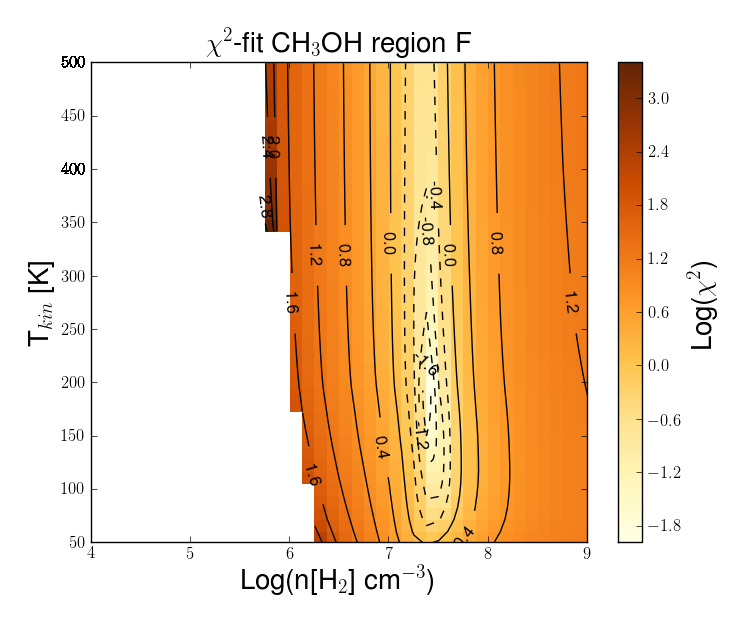}
     \caption{Pearsons $\chi^2$ results 
     of line ratios predicted by RADEX modelling and the observed ratio for all detected E-type CH$_3$OH lines of the three regions (\textit{Left}:A, \textit{Middle}:B, \textit{Right}:F). The color-scales are chosen such that lighter color indicates lower $\chi^2$ values and better agreement with observed line ratio and models. Densities are well fitted with values of 2$\times 10^7$ cm$^{-3}$ for region 'A', $>$3$\times 10^8$ cm$^{-3}$ for region 'B' and 4$\times 10^7$ cm$^{-3}$ for region 'F', }
     \label{fig:radexsma}
\end{figure*}
}

\def\PlaceFigureradexform{
\begin{figure}
\centering
\includegraphics[width=8cm]{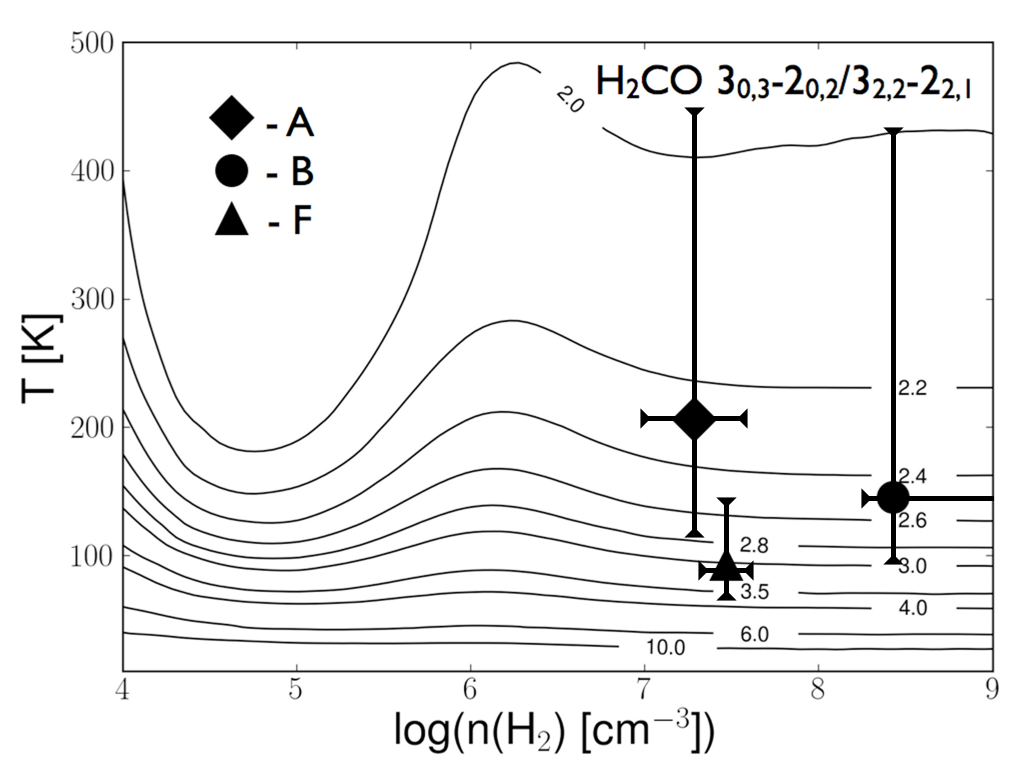}

     \caption{Line ratio of H$_2$CO 3$_{\rm{0,3}}$-2$_{\rm{0,2}}$ over 3$_{\rm{2,2}}$-2$_{\rm{2,1}}$ modelled with RADEX. Observed line ratios are shown with a diamond (A), circle (B) and triangle (F) at the best-fit densities derived using the CH$_3$OH line ratios (see text). Error bars for the uncertainties originate from the line ratio error (21$\%$). Errors on the densities originate from the $\chi^2$ fitting. Temperatures are 215 K (A), 150 K (B) and 100 K (F).}
     \label{fig:formal}
\end{figure}
}

\def\PlaceFigurewaterprofile{
\begin{figure}
\centering
\includegraphics[width=9cm]{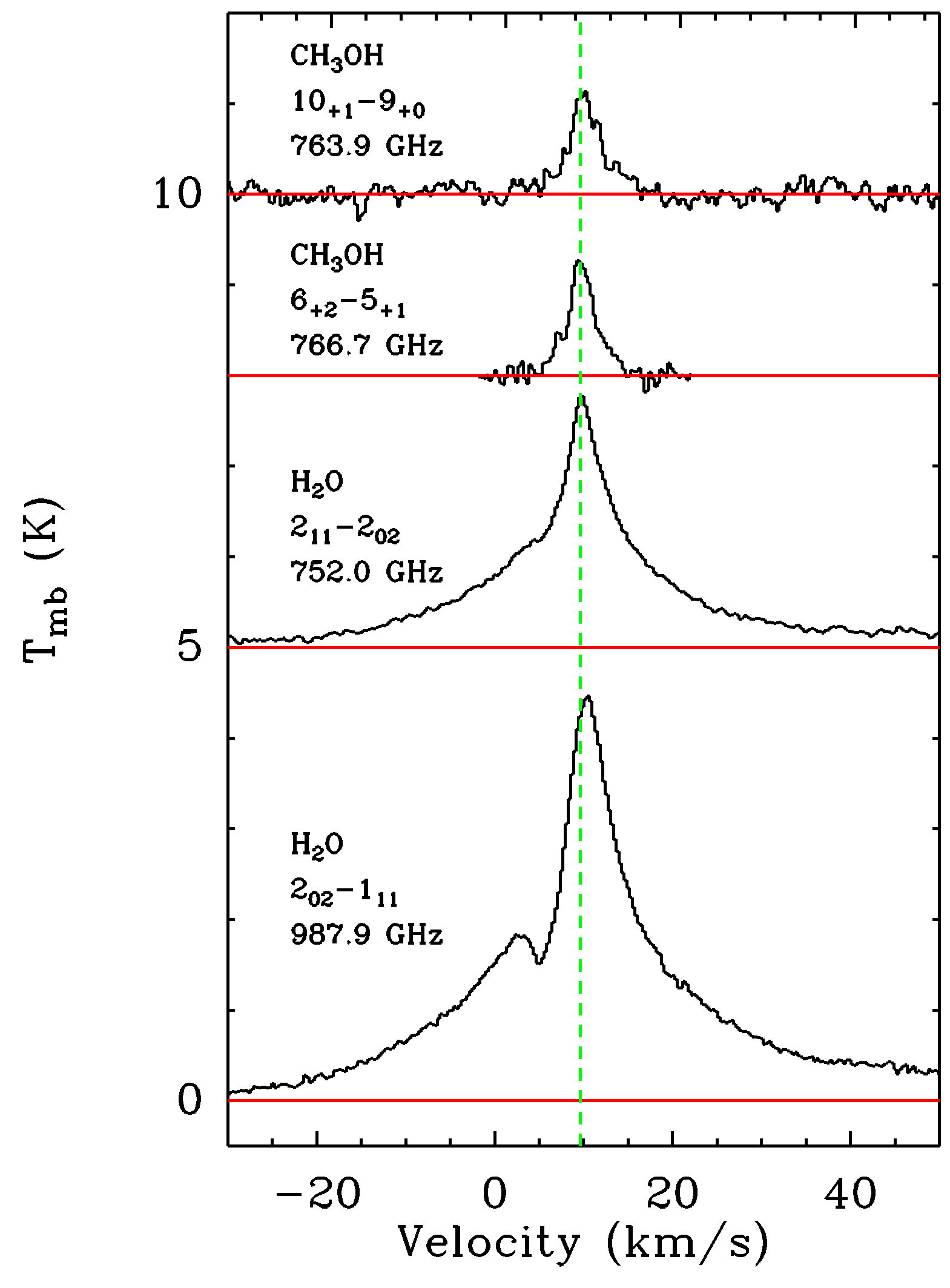}

     \caption{HIFI Line profiles of the CH$_3$OH 10$_{+1}$-9$_{+0}$ (763.9 GHz) and 6$_{+2}$-5$_{+1}$ (766.7 GHz) and H$_2$O 2$_{11}$-2$_{02}$ (752.0 GHz) and 2$_{02}$-1$_{11}$ (987.9 GHz) transitions. The CH$_3$OH lines are multiplied by a factor 5. Spectra are binned to 0.3 km s$^{-1}$ bins. Respective baselines are shown in red. The V$_{LSR}$ of NGC 2071 is shown in green. Note that the wings of the H$_2$O lines extent beyond the horizontal scale. }
     \label{fig:water}
\end{figure}
}

\def\PlaceFigurewaterratio{
\begin{figure}
\centering
\includegraphics[width=9cm]{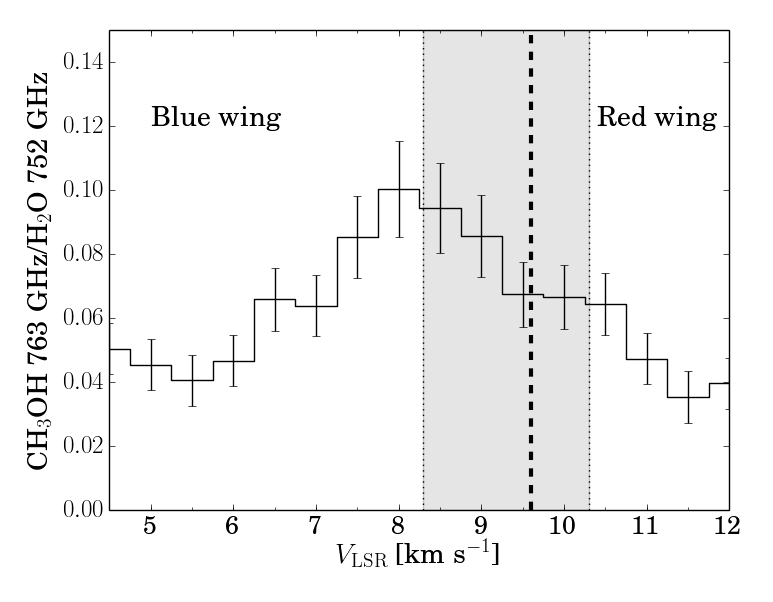}
     \caption{Line ratio of the CH$_3$OH 10$_{+1}$-9$_{+0}$ over the H$_2$O 2$_{11}$-2$_{02}$ transitions, both rebinned to 0.5 km s$^{-1}$. The $V_{\rm{LSR}}$ is shown with a dashed line. The range of velocities where the high optical depth of the water line is hypothesized to affect the derived ratio is shown in grey. No correction has been applied (see text).  }
     \label{fig:waterratio}
\end{figure}
}

%
\def\PlaceTableSettings{

\begin{table}
\caption{Observational parameters.}
\label{tab:set}
\begin{center}
\begin{tabular}{l c c c c c}
\hline \hline
\multicolumn{6}{c}{HIFI} \\ \hline
Setting & Freq.$^1$  & Band & $\theta_{mb}$ & ObsId. & RMS$^2$\\
 & GHz & & $''$ & & mK \\ \hline
1 & 547.676 & 1a & 38.0 & 	1342194490 & 5 \\
2 & 556.936 & 1a & 38.0 & 1342205274 & 15\\
3 & 752.033 & 2b & 28.0 & 1342194682 & 19\\
4 & 987.927 & 4a & 21.5 & 1342204503& 25\\  
5 &1095.67 & 4b & 19.0 & 1342227395	& 17\\
6 &1113.343 & 4b & 19.0 & 1342194790 & 18\\ 
7 &1153.127 & 5a & 18.5 & 1342206128 & 93\\ \hline
\multicolumn{6}{c}{SMA} \\ \hline
\multicolumn{3}{c}{Bandwidth} & \multicolumn{3}{c}{2$\times$4 GHz} \\
\multicolumn{3}{c}{Center Frequencies} & \multicolumn{3}{c}{230.5 \& 219.0 GHz} \\
\multicolumn{3}{c}{Configuration} & \multicolumn{3}{c}{SMA Compact} \\
\multicolumn{3}{c}{Beam size} & \multicolumn{3}{c}{3.4$''\times$2.9$''$} \\
\multicolumn{3}{c}{Observation Date} & \multicolumn{3}{c}{January 13th 2010} \\
\hline 
\end{tabular}
\end{center}
$^1$ Center Frequency of the HIFI Band.\\
$^2$ Measured in 0.5 km s$^{-1}$ bins.
\end{table}

}

\def\placehifilines{
\begin{table}
\caption{Emission lines detected with HIFI. The average uncertainty on the integrated intensity is estimated at 15$\%$ (see text)}.
\begin{center}
\begin{tabular}{lcccc}
\hline \hline
Transition & Type & Freq.  & E$_{\rm{up}}$& $\int T_{\rm{MB}} \rm{d}$$V$ \\ 
& &[GHz] & [K] & [K km s$^{-1}$] \\ \hline
 \multicolumn{4}{c}{CH$_3$OH} \\ \hline
 3$_{-3}$ - 3$_{-2}$ & E & 766.98 & 76.64 & 1.03 \\
 4$_{-3}$ - 4$_{-2}$ & E & 766.96 & 85.92 &  1.03 \\
 6$_{+2}$ - 5$_{+1}$ & A & 766.71 & 86.46 & 0.90 \\
 5$_{-3}$ - 5$_{-2}$ & E & 766.91 & 97.53 & 0.41 \\
 8$_{+0}$ - 7$_{+1}$ & E & 543.08 & 96.61 & 0.75 \\
 6$_{-3}$ - 6$_{-2}$ & E & 766.81 & 111.46 & 0.56 \\
 5$_{-4}$ - 4$_{-3}$ & E & 766.76 & 122.72 & 0.66 \\
 7$_{-3}$ - 7$_{-2}$ & E & 766.65 & 127.71 & 0.39 \\
 10$_{+1}$ - 9$_{+0}$ & A & 763.95 & 141.08 & 1.19 \\
 7$_{+4}$ - 6$_{+3}$ & A & 974.87 & 145.33 & 0.88 \\
 8$_{-3}$ - 8$_{-2}$ & E & 766.40 & 146.28 & 0.33 \\
 10$_{+2}$ - 9$_{+1}$ & A & 986.10 & 165.40 & 0.38 \\ 
 9$_{-3}$ - 9$_{-2}$ & A & 766.08 & 167.16 & $<$0.30 \\
 11$_{+1}$ - 10$_{+1}$ & A & 536.19 & 169.01 & 0.50 \\
 11$_{+2}$ - 10$_{+1}$ & E & 558.34 & 175.15 & 0.40 \\
 10$_{-3}$ - 10$_{-2}$ & E & 765.51 & 190.37 & 0.29 \\
 12$_{+1}$ - 11$_{+0}$ & E & 751.55 & 202.12 & 0.27 \\
 11$_{-3}$ - 11$_{-2}$ & E & 764.81 & 215.90 & 0.26 \\
 12$_{+3}$ - 11$_{+2}$ & E & 1109.58 & 228.78 & 0.33 \\
 15$_{+2}$ - 14$_{+1}$ & E & 754.22 & 300.98 & 0.36 \\
 18$_{+1}$ - 17$_{+0}$ & A & 1105.94 & 407.62 & 0.23 \\
 23$_{+1}$ - 22$_{-2}$ & A & 977.37 & 427.10 & 0.24 \\
 20$_{+0}$ - 19$_{-1}$ & A & 1095.06 & 497.93 & 0.23 \\ \hline
 \multicolumn{4}{c}{H$_2$O} \\ \hline
 2$_{02}$-1$_{11}$ & Para &  987.927 & 100.8 & 88.48\\
 2$_{11}$-2$_{02}$ & Para &752.033 & 136.9 & 46.62\\ 
\hline
\end{tabular} \\
\end{center}
\label{tab:hifi}
\end{table}
}

\def\placesmalines{
\begin{table*}
\caption{Detected CH$_3$OH, CH$_3$CN and H$_2$CO with the SMA. Lines are integrated over their width and converted to Kelvin km s$^{-1}$. Uncertainty of the flux density on all lines is 15$\%$}.
\begin{center}
\begin{tabular}{llclrrrr}
\hline \hline
Molecule & Transition & Type & Frequency  & $E_{\rm{up}}$& \multicolumn{3}{c}{$\int T_{\rm{MB}} \rm{d}$$V$} \\ 
& & & & & A & B & F \\
& & & [GHz] & [K] & \multicolumn{3}{c}{[K km s$^{-1}$]}\\ \hline
CH$_3$OH 
& 3$_{-2}$ - 4$_{-1}$   & E & 230.03 & 39.8  & 3.7 & 12.7 & 5.4 \\
& 4$_{+2}$ - 3$_{+1}$   & E & 218.44 & 45.5  & 5.7 & 37.1 & 50.5\\ 
& 8$_{-1}$ - 7$_{0}$    & E & 229.76 & 89.1  & 48.4& 64.2 & 76.7 \\
& 8$_{+0}$ - 7$_{+1}$   & E & 220.08 & 96.6  & 4.6 & 24.2 & 14.9 \\
& 10$_{+2}$ - 9$_{+3}$  & A & 231.28 & 110.9 & 8.2 & 14.0 & -  \\
& 10$_{+2}$ - 9$_{+3}$  & A & 231.42 & 110.9 & 19.4& 11.3 & 8.5  \\
H$_2$CO 
& 3$_{0,3}$ - 2$_{0,2}$ & para & 218.22 & 21.0 & 213 & 267.0 & 184.0\\
& 3$_{2,2}$ - 2$_{2,1}$ & para & 218.48 & 68.1 & 90.8 & 108.8 & 56.4\\
& 3$_{2,1}$ - 2$_{2,0}$ & para & 218.76 & 68.1 & 93.1 & 103.0& 53.9\\
CH$_3$CN 				 & ~~~0/1 & & 220.75 & 68.9 & 48.3 & 62.4& - \\
   			 				 & ~~~~2 & & 220.73 & 97.4 & 45.9 & 72.6& - \\
~~$J$ = 				 & ~~~~3 & & 220.71 & 133.2 & 26.0 & 34.2& - \\
 12--11					 & ~~~~4 & & 220.68 & 183.1 & 20.1 & 21.7& - \\
	ladder				 & ~~~~5 & & 220.64 & 247.4 &  - & 14.1& - \\
								 & ~~~~6 & & 220.59 & 325.9 &  - & 13.2& - \\
\hline 
\end{tabular} \\
\end{center}
\label{tab:sma}
\end{table*}
}

\section{Introduction}
Strong emission lines of hydrogenated carbon-bearing species are a common feature associated with the early stages of star formation. Embedded high-mass protostars possess a so-called `hot core' 
component \citep{2000prpl.conf..299K}, from which 
large quantities of emission lines from organic molecules have been detected \citep[e.g.,][]{2005ApJS..156..127C}. This spherically symmetric region around the massive protostar is powered solely by luminosities from the just-ignited protostar and its accretion flow.

Similar emission features of organic compounds
have been detected around low-mass protostars \citep{1995ApJ...447..760V,2003ApJ...593L..51C,2004ApJ...615..354B,2008A&A...488..959B,2011A&A...532A..23C}, including simple sugars \citep{2012ApJ...757L...4J}.  
Even though most emission lines of the organics clearly originate in the inner few hundred AU, their physical origin is fiercely debated. In one scenario, emission originates in a spherically symmetric component around the propostar similar to the hot core, referred to as the `hot corino' 
\citep{2007prpl.conf...47C}.
In some sources the molecular emission does not coincide with the positions where the continuum peaks and the protostar is believed to be located \citep[e.g.,][]{2005ApJ...632..371C}. Observed line shapes, in particular the line wings, do not support a spherical model. Furthermore, SED and envelope modelling of low-mass protostars show that although central warm regions are possible \citep{2002ApJ...575..337S}, typical sizes and column densities of the protostellar envelope are much larger and higher than predicted.
Other sources of energy are required to explain the observations.

Abundances of molecules such as methanol and formaldehyde (CH$_3$OH and H$_2$CO) are factors of $\sim$30-100 higher in protostellar environments than in dark clouds \citep{1998ApJ...501..723T}, where non-thermal desorption dominates \citep{2009A&A...494L..13O}. From examination of the physical and chemical conditions, release of 
icy grain mantles must play a key role \citep{2000A&A...361..327V,2005A&A...442..527M}. It is poorly understood in which situations thermal desorption of water\ (H$_2$O)
\citep{1993ApJ...417..815S} is dominant, when grain-grain collisions that shatter ice mantles
\citep{1996ApJ...469..740J} are required, or when sputtering by energetic particles produced in shocks play a large role \citep{2010MNRAS.406.1745F}. Whatever the role, CH$_3$OH is a pure grain mantle product.

\cite{2004ApJ...617L..69B} and \cite{2008A&A...488..959B} showed the importance of resolved interferometric observations of organic emission. By resolving components and their conditions,  differences in the formation of organics can be uncovered. Abundances of some species cannot be produced by grain surface hydrogenation alone. A combination of grain surface reactions and gas phase formation and/or destruction is often required.

Shock desorption has been invoked 
to account for observed offsets of organic emission when visibly associated with outflowing material 
\citep[e.g,][]{2002A&A...381...77B}. Shocks and their associated cavities also allow for additional energy injections not possible in spherical models.
Warm gas on the surface of outflow cavity walls has been invoked to explain observed CO $J$=6--5 emission \citep{1995ApJ...455L.167S,2009A&A...501..633V}. 
There is no reason why this cannot be a physical driver for emission of organics, following the relatively high gas temperature of this component \citep{2012A&A...537A..55V}. 
An important constraint to the conditions is the amount of CH$_3$OH destroyed, either through dissociative desorption during the sputtering or reactions with H in the shock. Recently, \cite{2014MNRAS.440.1844S} found that comparison of CH$_3$OH with H$_2$O can probe the conditions of irradiated shocks. 

Most likely, the various origins coexist. This is best shown by the ongoing work in intermediate mass protocluster OMC 2 FIRS 4 \citep{2010A&A...521L..39K,2013A&A...556A..57K,2013A&A...556A..62L}. The small scales show more than 4 components, while the observed methanol emission cannot be tied to one component, even though $>$ 100 lines have been detected. Multiple components and origins are required.

The tightly packed proto-cluster NGC 2071, located at 422 pc 
\citep{2012ApJ...746...71C,2012ApJ...751..137V} 
produces one of the most powerful outflows within 500 pc 
\citep{1986ApJ...303..416S,1990ApJ...364..164B}.
In combination with the relatively low
stellar masses of all three protostars and a cleaner environment, it is a better candidate to differentiate between
outflow driven scenarios and hot-core like origins or quantify the relative contributions of both. It is 
rich in molecular emission \citep{2003A&A...412..157J}, at times showing two gaussian components of different widths.

In this paper, we present new results using observations of emission 
lines of three organic compounds: methanol (CH$_3$OH), formaldehyde (H$_2$CO) and methyl cyanide (CH$_3$CN). These three organic compounds are commonly used as tracers for chemical acivity in nearby star 
forming regions \citep[e.g.,][]{2011A&A...534A.100J,2013A&A...556A..57K}.  Observations are  
obtained with ESA Herschel Space Observatory \citep{2010A&A...518L...1P} using the Heterodyne Instrument for the Far-Infrared (HIFI) \citep{2010A&A...518L...6D}, and the Sub-millimeter Array (SMA).
The aim of this paper is to determine the location, excitation and physical conditions 
of the region(s) responsible for the organic emission in NGC 2071 and 
differentiate 
between a hot corino scenario or one related to the outflowing gas. 
Section 2 presents the observations, and subsequent results are listed 
in section 3. The analysis is done in Section 4. The origin scenario is discussed in Section 5, with conclusions 
summarized in Section 6.

\PlaceFigureco
\PlaceTableSettings
\section{Observations}

Spectral line observations of HIFI \citep{2010A&A...518L...6D} and the
SMA were inspected for methanol (CH$_3$OH), 
formaldehyde (H$_2$CO) and/or methyl cyanide (CH$_3$CN) emission lines. 
To aid the analysis into the conditions that set the chemistry the 987 and 752 GHz H$_2$O lines are included in this paper \citep{2014MNRAS.440.1844S}. 

SMA observations of the line wings of $^{13}$CO $J$=2--1 (see Fig. \ref{fig:13co}) are used as a reference for the location  of the outflowing gas. Table \ref{tab:set} lists the observational settings of both the SMA and Herschel-HIFI observations. Data reduction was performed using a combination of the following software tools: the MIR package for IDL, MIRIAD, HIPE\footnote{HCSS / HSpot / HIPE is a joint development (are joint developments) by the Herschel Science Ground
Segment Consortium, consisting of ESA, the NASA Herschel Science Center, and the HIFI, PACS and
SPIRE consortia.} \citep{2010ASPC..434..139O}and CLASS in GILDAS\footnote{GILDAS is a software package developed by IRAM to reduce and analyze astronomical data; http://www.iram.fr/IRAMFR/GILDAS.}.

\PlaceFigureCH3CN
\PlaceFigurehifi
\PlaceFigurewaterprofile

\subsection{HIFI}
HIFI\footnote{HIFI has been designed and built by a consortium of 
institutes and university departments from across Europe, Canada and the 
United States under the leadership of SRON Netherlands Institute for Space
Research, Groningen, The Netherlands and with major contributions from 
Germany, France and the US. Consortium members are: Canada: CSA, 
U.Waterloo; France: CESR, LAB, LERMA, IRAM; Germany: KOSMA, 
MPIfR, MPS; Ireland, NUI Maynooth; Italy: ASI, IFSI-INAF, Osservatorio 
Astrofisico di Arcetri- INAF; Netherlands: SRON, TUD; Poland: CAMK, CBK; 
Spain: Observatorio Astron{\'o}mico Nacional (IGN), Centro de Astrobiolog{\'i}a 
(CSIC-INTA). Sweden: Chalmers University of Technology - MC2, RSS $\&$ 
GARD; Onsala Space Observatory; Swedish National Space Board, Stockholm 
University - Stockholm Observatory; Switzerland: ETH Zurich, FHNW; USA: 
Caltech, JPL, NHSC.} observations of NGC 2071 were carried out within the scope of the WISH key program \citep{2011PASP..123..138V}, targeting rotational emission of H$_2$O. All pointings were centered at a position with a Right Ascension of 05h47m04.4s and a Declination of 00d21m49s.
Except the setting targeting the ground-state H$_2$O line at 557 GHz (setting 2), all observations were done 
using DBS Fast Chop modes (See Table \ref{tab:set} for the observed HIFI bands and Observations Identification Numbers). 
After delivery, data were reprocessed with the default pipeline in HIPE 7.1 \citep{2010ASPC..434..139O} 
and calibration version HIFI$\_$CAL$\_$6$\_$0. Further data reduction was done using HIPE 8.1. 
H and V spectra were averaged together prior to converting the data to the velocity scale.   
The WBS spectrometer provided a 4 GHz bandwidth with a 1.1 MHz frequency resolution (0.7 to 0.4 km s$^{1}$). The beam of HIFI ranges from 18 to 41$''$ much larger than the structure seen in \cite{2012ApJ...751..137V} (see Table \ref{tab:set} and Fig. \ref{fig:13co}). Relative calibration errors are 15$\%$ for all bands except the one in Band 5, which is 20$\%$.

\subsection{SMA}
The Sub-millimeter Array (SMA) observed NGC 2071 in compact configuration on 
January 3rd 2010 using a bandwidth of 4 GHz centered at 230.538 GHz and 4 GHz centered at 219.634 GHz
and a spectral resolution of 0.3 kms s$^{-1}$\footnote{The calibrated data is available through the Radio Telescope Data Center hosted by the Smithsonian Astrophysical Observatory (http://www.cfa.harvard.edu/rtdc/}. The phase center was located at a Right Ascension of 05h47m04.7s  and a Declination of +00d21m44.0s, the same coordinates used to carry out the HIFI observations. Continuum results 
were published in \cite{2012ApJ...751..137V}. With a spatial resolution corresponding to 
$\sim$1300 AU (3.4$''\times$2.9$''$), the full extent of the flow is spatially resolved, as are the 
separations between most of the individual protostars \citep{2012ApJ...751..137V}. 
The noise levels were 4 mJy beam$^{-1}$ for the continuum and 0.08 Jy beam$^-1$ per 0.8 km s$^{-1}$ bin. this corresponds to $\sim$0.4 K in the 0.8 km s$^{-1}$ bins. 
Calibration uncertainty in the lines was derived to be 15$\%$ after the model fitting of Uranus. We will follow the notation from \cite{2012ApJ...751..137V}, in which the 
source IRS 3 is labelled as 'A', IRS 1 as 'B' and IRS 5 as 'C'. Note that the phase center is located in between 'A' and 'B'.
Due to the placement of correlator chunks, which optimized coverage for the CO isotopologues and continuum, 
several instrumental artifacts needed to be flagged
at semi-regular intervals. The only one affecting emission lines of H$_2$CO, CH$_3$OH or CH$_3$CN
appears between the $K$ = 1 and 2 transitions of the $J$ = 12--11 
ladder of CH$_3$CN. The artificial feature is 8 channels wide with a regular increase per 
channel. Spectra were corrected by fitting and sub-tracting a step-function to 
these channels. Figure \ref{fig:ch3cn} presents the uncorrected CH$_3$CN spectrum for source 'A'.

\placehifilines
\section{Results}

\subsection{HIFI}
In the HIFI bands, a total of 24 CH$_3$OH emission lines were identified, but no CH$_3$CN or H$_2$CO lines were seen. The H$_2$O lines are both very clearly detected with very broad line wings ($>$20 km s$^{-1}$, see Fig. \ref{fig:water}). The rms noise levels can be found in Table \ref{tab:set}. More details on water emission will be in a forthcoming paper (M$^{\rm{c}}$Coey \& van Kempen, in prep).
Quantum numbers, 
rest frequencies, energy levels and velocity integrated flux densities of the detected transitions are 
listed in Table \ref{tab:hifi}. Note that no lines were found in Band 5, which has a higher calibration uncertainty than the other Bands. As such the uncertainty for all lines in Table \ref{tab:hifi} is assumed to be 15$\%$.
Rest frequencies of all detections were checked against the molecular line survey of  \cite{2013A&A...556A..57K}. One feature was removed as questionable.
Energy levels range from 75 to 500 Kelvin. The bulk of the lines are observed 
around frequencies of 766 GHz, within the $K_{\rm{up}}$ to $K_{\rm{l}}$ = $-3$ to $-2$ ladder (See 
Fig. \ref{fig:ch3ohmethanol}).
All detected lines have $K_{\rm{up}}$ = $\leq|3|$.
Lines are consistently $\approx$6 km s$^{-1}$ wide. No evidence was found
for an increase of line width as a function of energy level, as suggested by \cite{2010A&A...521L..39K}. A very small shift in in velocity with respect to the source velocity is at times observed, but this is more likely to be an artifact of the methanol lines not being in the center of the HIFI bandpass and 
not a physical effect.

\placesmalines

\PlaceFiguremetha

\subsection{SMA}

Table \ref{tab:sma} lists the detected CH$_3$OH, H$_2$CO and CH$_3$CN transitions with quantum numbers, rest frequencies, energy levels and integrated flux densities. 
For CH$_3$OH, only transitions with $K_{\rm{up}} \leq |2|$ are seen, with 
one line $K_{\rm{up}}=0$. For CH$_3$CN, all detections are part of the $K$=12--11 ladder around 220.7 GHz.  All detected lines have line widths ranging between 5.9 to 8.0 km s$^{-1}$ wide,   
similar to the line widths of CH$_3$OH detected by HIFI.
Emission is spatially resolved into three positions for most lines (see Fig.
\ref{fig:maps}). Integrated flux densities are given for the three positions, labelled 'A', 'B' and 'F'. The 'F' position is a position 4$''$ west of B, with no continuum nor infra-red emission peak \citep{2012ApJ...751..137V}. 
Figure \ref{fig:13co} shows 'F' to be a position associated with entrained material. 
No velocity shifts were found between the three different positions.

The observed spatial distributions were fitted with Gaussian profiles with the goal to determine 
whether or not emission was spatially resolved. Spatial scales $>$1.5$\times$ the beam size along at least one axis were required. 
 It was found that only H$_2$CO emission from 'F' and 'B' is clearly spatially resolved. It is unresolved for 'A'. CH$_3$OH emission is predominantly unresolved at all three positions for all transitions. Only the CH$_3$OH transition at 218.44 GHz hints at resolved emission at 'F', albeit only marginally ($\sim 1.2\times$ the beam).  All CH$_3$CN emission is unresolved at positions 'A' and 'B' and undetected at 'F'. Note that SMA observations resolve out any structure larger than 13$''$ \citep{2012ApJ...751..137V}.
The narrow H$_2$CO component seen in \cite{2003A&A...412..157J} between 2.2 and 2.9 km s$^{-1}$, is not detected and assumed to be resolved out. It thus originates within a component that is smoothly distributed on scales of $\sim$4,000 AU or larger. 

After correcting for the difference in beam size, the sum of the three regions seen in the H$_2$CO 3$_{03}$-2$_{02}$ transition reproduces 96$\%$ of the emission listed in \cite{2003A&A...412..157J}. 
For the H$_2$CO 3$_{22}$-2$_{21}$ transition, the 16 km s$^{-1}$ wide component is not detected by the SMA. The sum of the three regions equals to 75$\%$ of the total intensity reported in \cite{2003A&A...412..157J}. It should be noted that the wide velocity component seen in \cite{2003A&A...412..157J} was solely detected in the H$_2$CO 3$_{22}$-2$_{21}$ transition, and not in any other line included in their study.
For the 4$_2$-3$_1$ line at 218.44 GHz, the three regions reproduce 91$\%$ of the observed JCMT emission.
Given the uncertainties in absolute flux calibration of both the JCMT (20$\%$) and SMA (15$\%$),  organic emission seen with the JCMT is almost fully recovered by the SMA by summing the emission of the three regions.
We thus assume that for all lines detected by the SMA, emission is produced in these three regions and not in any undetected component resolved out, or other regions within the JCMT beam.

\PlaceFiguresmaradex
\section{Analysis}
\subsection{Excitation of CH$_3$OH and H$_2$CO}
Most CH$_3$OH and CH$_3$CN transitions inherently possess high critical densities \citep[10$^9$ cm$^{-3}$ or higher, see][]{2003A&A...412..157J}. 
As such, excitation conditions can reliably be derived using the 1-D non-LTE radiative transfer code RADEX \citep{2007A&A...468..627V} from line ratios. The transitions are assumed to be completely optically thin and fill the beam, following 
\cite{2010A&A...521L..39K}. Optical depth effects and other radiative transfer effects such as IR pumping are discussed later.

A large grid of RADEX models was run to derive E-type CH$_3$OH, H$_2$CO and CH$_3$CN emission line predictions between 215 
and 1115 GHz, using datafiles provided by LAMDA\footnote{the Leiden Atomic and Molecular DAtabase: See http://www.strw.leidenuniv.nl/~moldata/}
 \citep{2005A&A...432..369S} with the upward rate coefficients calculated at the appropriate temperature. 
Temperatures ranged from 50 to 500 Kelvin, while densities were varied between 10$^4$ cm$^{-3}$ and 10$^9$ cm$^{-3}$. 
 
Most collisional rate coefficient are limited to  $J_{\rm{up}}$ = 15 and temperatures of 200 K. At higher 
kinetic temperatures, downward collisional rate coefficients of 200 K are used. 
The RADEX line width was set to 6 km s$^{-1}$. A column density of only 10$^7$ cm$^{-2}$ was chosen, since we are interested in line ratios of optically thin transitions. Differential beam dilution is corrected using the method described in the appendix of B of \citep{2010A&A...522A..91T} assuming the regions
are point sources. Dilution factors are thus proportional to $\theta_{MB}^{-2}$. However, test using the one-dimensional source structure ($\theta_{MB}^{-1}$) show little to no changes. 
 
Predictions of line ratios were compared to the observed ratios of the five E-type CH$_3$OH lines detected with the SMA for each region. 
Figure \ref{fig:radexsma} shows the results of a Pearson's $\chi^2$ test\footnote{Pearson's test is defined as $\chi^2 = \sum_{i=1}^{n} \frac{(R_i - P_i)^2}{P_i}$) with $R_i$ the observed ratio  and $P_i$ the calculated ratio over a total of $n$ ratios, in which the smallest $\chi^2$ gives the best fit.}. The errors on individual ratios are 21$\%$.
The ratios reveal clear differences in the physical conditions between the regions, presented in Fig. \ref{fig:radexsma}. 
The best fit found for region 'F' is a density of 4$\pm 0.1\times$10$^7$  cm$^{-3}$. 
Region 'A' is constrained to a density of 2$\pm 0.8 \times$10$^7$ cm$^{-3}$, while region 'B' is constrained to  densities $>$ 2$\times$10$^8$ cm$^{-3}$. 
Temperatures are loosely constrained to values $>$ 100 Kelvin. 

Better constraints on the kinetic temperature can be obtained through comparison of the line ratio of H$_2$CO  3$_{\rm{0,3}}$-2$_{\rm{0,2}}$ over 3$_{\rm{2,2}}$-2$_{\rm{2,1}}$ with the RADEX model. 
Figure \ref{fig:formal} shows the resulting line ratios of our RADEX grid of these two  H$_2$CO lines.
The observed ratios of the three regions are plotted at the densities derived using the $\chi^2$ tests of the CH$_3$OH ratios. This results in temperature constraints of 215$^{+200}_{-70}$ K for region 'A', 150$^{+280}_{-50}$ K for region 'B' and 100$^{+40}_{-15}$ K for region 'F'. The error bar on the temperature can be significant, especially for regions 'A' and 'B'. Much higher temperatures ($>$ 400 K) are within uncertainty. This is in part due to the lack of collision rate coefficients above 300 K for H$_2$CO. However, constraints on lower temperatures are much better. Even with these large error, all three regions are warmer than the surrounding envelope, but not hot ($T > $ 500 K).
Due to the molecular structure of H$_2$CO, there is no difference between using the 3$_{\rm{2,2}}$-2$_{\rm{2,1}}$ or 3$_{\rm{2,1}}$-2$_{\rm{2,0}}$ transition. Given the near identical line ratios of either with respect to the 3$_{\rm{0,3}}$-2$_{\rm{0,2}}$ line, the temperature constraint can be assumed to be robust and not affected by observational biases.

The narrow range of constraints for 'F' imply that the emission in that region comes from a more homogeneous medium than 'A' or 'B'. The less restrictive fits in density around the protostellar positions 'A' and 'B'  likely indicate density gradients within the emitting region. 
The significant differences in density (2$\times$10$^7$ cm$^{-3}$ vs. 3$\times$10$^8$ cm$^{-3}$) and, to a lesser extent, temperature (215 K vs. 150 K) between 'A' and 'B' are surprising
given the lack of difference in line width (6 km s$^{-1}$ for both), individual line emission of the H$_2$CO  3$_{\rm{2,2}}$-2$_{\rm{2,1}}$ line (109 vs 91 K km s$^{-1}$), 3$_{\rm{0,3}}$-2$_{\rm{0,2}}$ line (267 vs 213 K km s$^{-1}$),  resolved dust emission \citep[0.13 Jy beam$^{-1}$ vs. 0.12 Jy beam$^{-1}$, see Table 3 and Fig. 2 in][]{2012ApJ...751..137V} or stellar (0.9 \Msol vs. 0.5 \Msol) and individual envelope masses (8.2 \Msol vs. 14.2 \Msol). The main difference between the two regions is that 'B' powers the large NGC 2071 outflow, while 'A' possesses a much weaker flow \citep{2012ApJ...746...71C}.

\subsection{Optical depth}
The assumption that the observed methanol lines are all optically thin must be investigated further. Using RADEX one can determine the column densities and in turn the associated optical depths. It should be noted that the solution is degenerate with an assumed source size within a single beam. Smaller source sizes require higher column densities and thus higher optical depths are to reproduce the same amount of emission. 
If the source fills the SMA beam, as assumed above, a column of 4$\times$10$^{15}$ cm$^{-2}$ was derived to best fit the observed flux densities. In this model, optical depths are typically 0.01 or lower. 
RADEX shows that to produce optically thick lines with the observed emission, a column of 8$\times$10$^{17}$ cm$^{-2}$ or higher is required. Such a column corresponds to a source size 14 times smaller than the SMA beam, equal to $\sim$90 AU. 

However, at these high column densities, some lines will mase, affecting line ratios. The 218.44 GHz line is a weak maser, but shows similar line strengths when masing. The largest effect is seen for the ratios between the 218.44, 220.08 and 229.76 GHz lines. At higher columns, RADEX shows the 229.76 line as an absorption feature of similar depth as the emission of the 218.44 GHz line. However, no absorption was seen. In addition, the 220.08 line should be 100 times brighter than the 218.44 GHz line, while a ratio of about 0.5 to 2.5 was observed. From this, we can thus exclude that the emission originates from a very small region with a very high column density.
 The theoretically largest column that correctly reproduces the observed line ratios is a few times 10$^{16}$ cm$^-2$, corresponding to an optical depth ($\tau$) of 0.4. The assumption that all CH$_3$OH emission is optically thin is thus justified.

Besides a high optical depth, IR pumping of low density (10$^3$ cm$^{-3}$) methanol gas  could also influence the observed line ratios for $v$=0 rotationally excited transitions \citep{2007A&A...466..215L}. Observations of torsionally excited methanol lines ($v$=1) are typically invoked to break the degeneracy. Although over 6 torsionally excited transitions were covered in the bandpass of the SMA, none were detected for the achieved sensitivity of (71 mJy in 2 km s$^{-1}$ bins). In addition, no $v$=1 transitions were detected in the HIFI bandpass. This indicates IR pumping has little to no effect on the methanol emission.
RADEX experiments with a grey body radiation field and lower densities showed that although the ratios of non-masing lines are indeed reproduced, it is impossible to simultaneously quench all three maser lines and produce the correct ratios. 
It is possible to add a very small ($<$50 AU) IR pumped component to the observed ratios. This can account for 25$\%$ of the observed line strengths. Whether or not such a component is present requires deep observations of the $v$=1 transitions, e.g. using the band around 241.2 GHz.

\PlaceFigureradexform
\PlaceFigurewaterratio
\subsection{Excitation of CH$_3$CN}

RADEX  results for CH$_3$CN line ratios cannot improve on the solutions for the physical conditions obtained with H$_2$CO and CH$_3$OH, as all detected lines are part of the $J$=12--11 ladder. 
From the RADEX models we derive a kinetic temperature of 200 K or higher for 'A' and $\sim$150 K for 'B'.  These values are consistent with the results of CH$_3$OH and H$_2$CO. It is thus reasonable to assume that the
CH$_3$CN emission originates from the same gas as CH$_3$OH and H$_2$CO in 'A' and 'B'. However, CH$_3$CN emission is absent from 'F'.

With the optically thin assumption, the similarity of line ratios between CH$_3$CN, H$_2$CO and CH$_3$OH in 'A' and 'B', and the observed strength of the emission of H$_2$CO and CH$_3$OH in 'F', 
the relative abundance of CH$_3$CN w.r.t. H$_2$CO and CH$_3$OH must be almost two orders of magnitude lower at 'F' then in 'A' and 'B'. Excitation conditions cannot explain the lack of CH$_3$CN emission at 'F'.

\subsection{Velocity profile: comparison to H$_2$O}
The behaviour of the observed line ratio between CH$_3$OH and H$_2$O emission lines away from the source velocity differentiates between two potential formation routes of H$_2$O \citep{2014MNRAS.440.1844S}.
If gas-phase CH$_3$OH and H$_2$O are created solely through grain mantle evaporation, the ratio will be flat. However, if gas-phase synthesis of H$_2$O is taking place, the ratio will drop as more H$_2$O is created from shocked material. A significant optical depth in the water line causes  observed ratios to be higher than physical processes would produce.
Although an optical depth cannot be derived due to a lack of observed isotopologues, it is hypothesized that the H$_2$O 752 GHz may be affected near the line center (between 8.3 to 10.3 km s$^{-1}$). In outflows of low-mass protostars \citep[e.g.,][]{2010A&A...521L..30K}, optical depth are factors of a few at line center.

However, due to the high critical density of water transitions ($>$10$^8$ cm$^{-3}$ for both the 987 and 752 GHz line), the assumed optically thick emission is most often effectively thin.
The effects of water opacity on the ratio are still relatively small. See e.g., Fig. 3 of \cite{2014MNRAS.440.1844S}. For velocities where optical depth may play a role,the ratio is overestimated by a factor 2 at most.

Figure \ref{fig:waterratio} shows the ratio of 763 10$_1$-9$_0$ CH$_3$OH transition and the 752 2$_{11}$-2$_{02}$ H$_2$O line. The observed CH$_3$OH/H$_2$O ratio clearly drops as a function of velocity on the blue wing. The red wing also is falling, although the number of channels not affected by optical depth is marginal. Velocities between 8.5 and 10.2 km s$^{-1}$ are thought to be affected by optical depth. As seen in the lower excited 987 GHz H$_2$O line, optical depth is expected to be at its highest there causing the ratios at these velocities to be higher.

Line ratios of other CH$_3$OH/H$_2$O combinations (e.g., the group of methanol lines at 766 GHz) show near identical profiles: A clear drop in the ratio on the blue side in combination with a drop on the red side and a middle likely affected by optical depth. Ratios derived using the 987 GHz H$_2$O line are more affected by optical depth, but still reproduce the falling ratio of the blue wing.

\section{Discussion: Origin of organic emission.}

Using the optically thin approximation, the observed spatial separation between the regions and lack of large-scale CH$_3$OH emission, CH$_3$OH HIFI detection originate solely in the three regions. The detected flux density for each line is a simple sum:
\begin{equation}
CH_3OH = C*(X_A*R_A + X_B*R_B+X_F*R_F)
\end{equation}
$CH_3OH$ is the total observed emission within the HIFI beam. $R_X$ is the predicted emission from RADEX of region 'A', 'B' or 'F' for that transition, and X the relative contribution. C is a constant to ensure the sum 
equals observed intensities.

If it is assumed $X_A$, $X_B$ and $X_F$ do not vary for individual CH$_3$OH lines, and are thus not vectors themselves, the set of equations can be solved by rewriting the equation for individual CH$_3$OH line contributions into a vector notation covering all the HIFI CH$_3$OH lines.  
\begin{equation}
\vv{CH_3OH} = \vv{C}*(X_A*\vv{R_A}+X_B*\vv{R_B}+X_F*\vv{R_F})
\end{equation} 
$\vv{CH_3OH}$ is a vector, one column wide, containing the HIFI-detected emission line flux densities. $\vv{R_X}$ are the vectors, again one column wide, containing the predicted fluxes obtained from RADEX for the 14 HIFI E-Type methanol transitions with $J$ $\leq$15. Higher $J$ transitions are not available in the molecular datafile. A-Type methanol is not included, since no constraints from the SMA are given for A-Type methanol transitions. Even assuming they originate under identical conditions with an abundance ratio of 1:1, inclusion of A-Type methanol does not contribute to the analysis below.  $\vv{C}$ is the vector containing the constant $C$ for each transition. Under the assumption that the source size is the same for all transitions, $\vv{C}$ will depend on the model parameters. Its only dependency is on the frequency of the transition. This known quantity  determines the beam dilution in the HIFI beam.
We realize that Eq. 2 is an oversimplification, as $X_A$, $X_B$ and $X_F$ may vary from transition to transition. Similarly, source sizes may become smaller for higher excited transitions. However, given the uncertainties in the observed emission, these simplifications produce acceptable results.   \\
Solving Eq. 2 using a least squares method, optimal values of 13$\%$ for $X_A$, 71$\%$ for $X_B$ and 16$\%$ for $X_F$ are derived. With constant calibration uncertainties (in the case for Herschel a conservative value of 10$\%$ on the flux densities is assumed), the square of the error on the relative contributions equals the mean error times the rank of the matrix divided by the number of transitions
\footnote{The error $\sigma$ is calculated from $\sigma^2 = \sigma_{f}^2*M/N$ with $\sigma_{f}$ the mean flux density error, M the rank of the matrix and N the number of entries.}. The solution actually has a rank of only 2 (as opposed to 3 as was expected). The reason for this reduction in rank is that emission contributions in the conditions for 'F' and 'B' are very similar for the transitions probed. These produce similar ladders and none of the transitions used can distinguish them. The statistical error on the values of 13, 71 and 16 $\%$ is 4$\%$.

For the detected SMA transitions, the mean of the individual relative contributions is indeed close to this value. Relative contributions of some of the individual transtions also adhere to the relation. For example, the CH$_3$OH 3$_{-2}$-4$_{-1}$ line in Table 3, 'A' contributes 17$\%$, 'B' contributes 58$\%$ and 'F' contributes 25 $\%$.
However, clear variation are visible. Stronger contributions from 'F' are detected for some transitions. Whether or not these are introduced by optical depths effects due to the beam at 'B', or if the inferred density gradient at 'B' comes into play, remains to be seen. 

Tests using other density and temperature parameters for regions 'A' and 'B' (i.e., higher temperatures and densities) also show variations at the level of 5$\%$ in the inferred contributions.

Deviations in relative contributions for higher excitation lines were subsequently investigated. Changes larger than 20$\%$  in relative contributions to higher excitation lines (e.g. by forcing 'A' to contribute 50$\%$ of the observed emission of a CH$_3$OH line with $E_{\rm{up}}$= 400 K) become very problematic. In such cases, HIFI detections with $E_{\rm{up}}$ $<$ 250 K significantly violate the constraints set by the SMA detections.
Variations of 5-10$\%$ are likely from transition to transition. 
In addition, emission from 'B' and 'F' cannot be disentangled from Herschel transitions alone. The percentages of 'B' and 'F' originate from the best solution in the least squares fitting and the mean of the SMA emission quantities. 

The absence of molecular data beyond a $J$ = 15 likely hinders further interpretation. To improve on this analysis, a full radiative transfer model with power law exponents describing envelope structures is warranted. This is considered to be beyond the scope of this paper. \\

In the end, the emission contribution from each region to all HIFI detections are estimated to be :
\begin{itemize} 
\item 15-20$\%$ originates in region 'A'
\item 60-70$\%$ in region 'B' 
\item 15-25$\%$ in region 'F'.
\end{itemize}

In region 'F', it is clear that the emission is related to the outflowing gas in position, but not in velocity. The absence of any stellar source rules out any possibility of disk-related scenarios or a hot corino. Whether or not the emission from 'A' and 'B' are related to the outflow or are part of a hot corino is debatable. However, a case in which the outflow plays a strong role is more likely.

Typical spherical envelope models with power-law density profiles show the derived higher densities of a few times 10$^7$ cm$^{-3}$ to a few times 10$^8$ cm$^{-3}$ to be located on scales of $\simeq$100 AU.
Gas temperatures of 150-200 K are located at scales of 50 AU or less, very close to the inner edge of the model \citep{2002ApJ...575..337S,2002A&A...389..908J,2012A&A...542A...8K}.
Envelopes around the individual NGC 2071 sources have been found to be very similar to these typical low-mass protostellar envelopes \citep{2012ApJ...751..137V}.

A scenario in which organics formed within the icy grain mantles, were subsequently liberated by either sputtering or grain-grain collisions and are currently emitting in a compressed, UV heated gas component is much more likely. 
Recently, \cite{2013A&A...557A..23K} found kinematic components in H$_2$O resolved lines consistent with  dissociative jet shocks in the inner 100 AU to explain distinct velocity components in H$_2$O spectra. 
Such components produce emission offset from the central velocity and as such cannot originate in a hot-core like environment. They are too powerful to be associated with potential non-dissociative shocks along the outflow cavity wall. An irradiated shock was inferred from the low abundances in that component. 

Outflows are known to be present at the smallest scales: 'B' is known to be 
the launching point of the large scale bipolar jet of NGC 2071 \citep{2009ApJ...701..710S}.
'A' is a source of strong water masers. The maser spots display significant velocity shifts, indicative of jet activity \citep{1998ApJ...505..756T,2012ApJ...746...71C}.

\subsection{Evidence for CH$_3$OH destruction and H$_2$O gas-phase synthesis}
The decreasing ratio of the line profiles (see Fig. \ref{fig:waterratio}) shows that the chemistry in NGC 2071 is likely similar as that found for NGC 1333 IRAS 4A and 4B \citep{2014MNRAS.440.1844S}. Atomic hydrogen formed in the shock is able to destroy CH$_3$OH through chemical reactions featuring hydrogen extraction, e.g. H + CH$_3$OH $\rightarrow$ H$_2$ + CH$_2$OH or H + CH$_3$OH $\rightarrow$ H$_2$ + CH$_3$O \citep{2005JPCRD..34..757B}. Both products are also quickly destroyed.  All these reactions have activation energies of $\sim$800 K, much lower than similar reactions capable of destroying H$_2$O (activation energy of $\sim$10$^4$ K).
Gas-phase synthesis of H$_2$O apparently can take place along the entire flow, given the brightness of the lines. Methanol emission is constrained to regions 'A', 'B' and 'F' by local physical conditions.
Only regions of sufficient density (10$^7$ cm$^{-3}$) can shield methanol from the high-temperature gas and subsequent dissociation.
The high spatial resolution observations (Fig. 5) show that the regions where these physical conditions occur are small ($<$1500 AU in size).  Observationally, we can conclude that the most likely location of these regions is in the neighborhood of protostars. This is not surprising considering envelope densities. However, correct conditions can also  occur elsewhere, with no protostar nearby (region 'F').

\subsection{Evidence for ice processing of CH$_3$CN}
The main difference between CH$_3$CN and CH$_3$OH is its formation. Both form predominantly 
on grain surfaces, but their reactions proceed differently under cold conditions.
CH$_3$CN is mainly formed through CN reacting with CO and other carbon containing species, while CH$_3$OH is a hydrogenation product of  CO. At low temperatures, hydrogen moves freely.
However, due to the low mobility of CN, CH$_3$CN formation requires heating and/or irradiation. \cite{2008ApJ...682..283G} show an increase of more than order of magnitude in both gas and ice abundances for CH$_3$CN as a function of time in various warm-up scenarios, while the CH$_3$OH or H$_2$CO formation remains unaltered compared to the cold phase formation. Additional gas-phase formation routes for CH$_3$CN after grain mantle evaporation cannot be invoked, as these typically produce abundances 
almost two orders of magnitude lower than the grain surface formation route \citep[see Fig. 7 to 9 and appendix 2 of][]{2008ApJ...682..283G}. 

At region 'F', little to no ice processing must have taken place due to the absence irradiation/heating at earlier stages, as opposed to the presence of a protostar at 'A' and 'B'. The young dynamical age of the NGC 2071 outflow ($<10^4$ yr, van Kempen et al., in prep) further corroborates this, limiting the production of CH$_3$CN.

\section{Conclusions}
This paper presented new data obtained with the SMA and Herschel-HIFI on the emission of organics around the cluster NGC 2071 and its outflow. The data were analyzed using a non-LTE radiative transfer code. The following conclusions are drawn:
\begin{itemize}
\item All organic emission detected by HIFI and the SMA observations originates in two or three regions, labelled 'A', 'B' and 'F'. 'A' and 'B' are  the positions of the two central protostars, while 'F' is a shock position in the blue outflow. 'F' is spatially resolved in some lower excitation lines, but does not emit in CH$_3$CN.
\item Using the non-LTE radiative transfer code RADEX, physical conditions of the three individual regions were constrained using a $\chi^2$ analysis of the H$_2$CO and CH$_3$OH emission. Densities higher than 10$^7$ cm$^{-3}$ (10$^8$ cm$^{-3}$ in the case of 'B') and temperatures of 100 K or higher were derived.
\item All HIFI CH$_3$OH detected are a sum of the emission of the three regions. On average 60-70$\%$ originates from 'B', 15-20$\%$ from 'A' and 15$\%$-25$\%$ from 'F'. No line originates solely from either 'B' (the densest region) or 'A' (the hottest region).
\item The high densities ($>10^7$ cm$^{-3}$) of the regions shield the gas-phase CH$_3$OH. 
Everywhere else, CH$_3$OH is destroyed through high-temperature reactions with hydrogen and/or through destructive sputtering.
\item The significant difference between CH$_3$OH and H$_2$O line wing shapes and extent suggests H$_2$O formation through gas-phase synthesis along the flow.
\item The lack of CH$_3$CN emission at 'F' is attributed to a difference in ice processing history.

\end{itemize}
Small-scale structure of organic emission is of vital importance in determining its origin and thus understand chemical processes during the early stages of star formations. Future observations with ALMA such as done in \cite{2013ApJ...779L..22J}  are needed to determine if all organics emission lines originate from a shock-induced scenario or a hot corino. In the case of NGC 2071, the complexity of the chemistry could not be understood by single-dish observations alone.

\begin{acknowledgements}
The research of Tim van Kempen is made possible by the Allegro ARC node. Michel Fich, Carolyn M$^{\rm{c}}$Coey and Sam Tisi are supported in this work by a Discovery Grant from NSERC and a Space Science Enhancement Program grant from the Canadian Space Agency.
Doug Johnstone is supported by the National Research Council of Canada and by a Natural Sciences and Engineering Research Council of Canada (NSERC) Discovery Grant. We thank Mihkel Kama for useful discussions on methanol detections using HIFI. Michiel Hogerheijde and Floris van der Tak are thanked for assistance on RADEX. The WISH team, in particular Silvia Leurini, is thanked for usefull discussions.

\end{acknowledgements}

\bibliographystyle{aa} 
\bibliography{2071_chem}

\end{document}